\documentclass[showpacs,twocolumn,preprintnumbers,amsmath,amssymb,pra]{revtex4-2}

\usepackage{graphicx}
\usepackage{amsmath}
\usepackage[colorlinks=true, allcolors=blue]{hyperref}
\usepackage{braket}

\begin{document}

\title{
 Gain-induced spectral non-degeneracy in type-II parametric down-conversion
}

\author{Behnood Taheri$^{1,3}$}
\email{behnood.taheri@uni-paderborn.de }
\author{Denis Kopylov$^{2,3}$} 
\author{Manfred Hammer$^{1,3}$}
\author{Torsten Meier$^{2,3}$}
\author{Jens Förstner$^{1,3}$}
\author{Polina Sharapova$^{2}$}
\affiliation{
$^{1}$Theoretical Electrical Engineering, Paderborn University, Paderborn, Germany.\\
$^{2}$Department of Physics, Paderborn University, Paderborn, Germany.\\
$^{3}$Institute for Photonics Quantum Systems (PhoQS), Paderborn University, Paderborn, Germany.
}

\date{\today}

\begin{abstract}
We demonstrate the novel effect of gain-induced spectral shifts in the type-II parametric down-conversion (PDC) process, which results in a transition from degenerate to non-degenerate PDC with increasing parametric gain. This effect, originating from the second-order dispersion terms, significantly alters the properties of PDC in the high-gain regime, where it leads to increased distinguishability of the generated photon pairs. The effect is established by evaluating a rigorous theoretical model, which is based on solving a system of coupled integro-differential equations for monochromatic operators. The widely used spatially-averaged approximate model fails to reproduce this important effect.
\end{abstract}

\maketitle

\paragraph*{Introduction.}
Squeezed states of light are a key ingredient of continuous variable quantum computing \cite{Weedbrook,Zhong_phase,Madsen2022}, quantum communication \cite{Braunstein-book}, and quantum sensing \cite{LIGO,Schnabel_2017}. Currently, one of the most flexible and efficient platforms for their generation is based on the pulsed single-path high-gain parametric down-conversion (PDC) in waveguides \cite{Eckstein_2011,CHEKHOVA201527,Lemieux_2016}. Tunable waveguide dispersion, strong field confinement, together with high-repetition pump rates, make waveguide-based quantum light sources a promising framework for integrated quantum optical circuits~\cite{Kai_2019, Nehra_2022, Williams_2025}. 

To be efficient, continuous variable quantum technologies require high levels of quantum mode squeezing \cite{Yukawa2008,Zhang2006} and small sizes of squeezed light sources \cite{Dutt2015,Vernon2019,Lenzini2018}, making the generation of high-gain PDC in waveguides particularly interesting. However, if the gain is high enough, causality, which is represented by time-ordered (or spatially-ordered) exponentials in the quantum-mechanical description~\cite{Christ_2013,Quesada_2014,Horoshko_2019},  begins to play a significant role in the generation process. Moreover, in the presence of strong material dispersion, the pulse propagation inside the waveguide is modified; therefore, it is necessary to consider the full dynamics of interacting fields. Theoretical models that incorporate spatial-ordering cannot be treated entirely analytically: only a numerical solution provides a rigorous description ~\cite{Sharapova_2020,Quesada_2022,Thekkadath_2024,Kopylov_2025}. The existing approximations like  narrowband-~\cite{Dayan_2007_theory} and plane-wave-pump~\cite{klyshko1988book,Kolobov1999} approximations, or spatially-averaged models ~\cite{Wasilewski_2006,Sharapova_2015} result in simple semi-analytical solutions; however, they do not capture the full dynamics and, as a consequence, may give misleading results for high-gain pulsed regimes. 

\begin{figure}[h]
    \includegraphics[width=\linewidth]{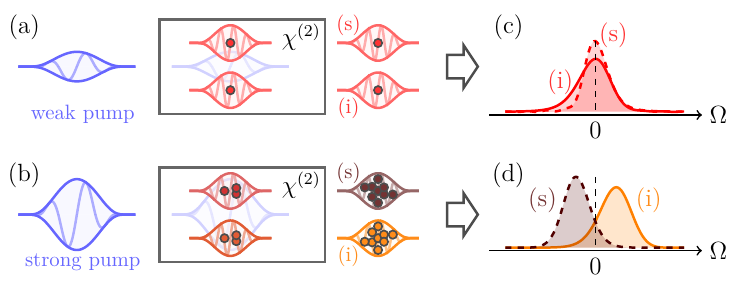} 
    \caption{The scheme of the (a)  low-gain and (b) high-gain regime of type-II PDC. Signal (s) and idler (i) fields belong to the orthogonal polarizations. (c,d) The spectra of the generated signal (dashed) and idler (solid) fields for (c) low and (d) high gain, respectively.
    }
\label{fig_1_scheme}
\end{figure}

In this letter, we study a new regime of pulsed single-path high-gain type-II PDC in highly dispersive waveguides, taking  spatial-ordering effects fully into account.  We show that, depending on the waveguide dispersion, the spectrum of high-gain PDC can differ significantly from its low-gain counterpart. In addition to the well-known effect of spectral broadening  \cite{Spasibko_2012,Christ_2013}, we demonstrate a new effect of  \textit{gain-induced non-degeneracy}, where the central frequencies of the signal and idler fields shift relative to their degenerate positions as the gain increases, providing non-overlapping spectra at high gain (Fig.~\ref{fig_1_scheme}). We demonstrate that this effect cannot be accurately predicted by theories that neglect spatial-ordering or high-order dispersion terms.  

\paragraph*{Theoretical model of high-gain type-II PDC.}
To describe high-gain PDC, we use the Rigorous Model (RM) \cite{Christ_2013}, which is based on the spatial propagator 
\begin{equation}
    \hat{G}(z) = \Gamma \iint d\omega d\omega^\prime J(\omega,\omega^\prime,z) \hat{a}^\dagger(\omega) \hat{b}^\dagger(\omega^\prime)  + h.c.
    \label{eq_generator_full}
\end{equation}

Here, $J(\omega, \omega^\prime, z) =  S(\omega + \omega^\prime) e^{i\Delta k(\omega, \omega^\prime) z}$ is the $z$-dependent coupling function that depends on the pump spectrum $S(\omega+ \omega^\prime)$ and the wavevector mismatch defined as $\Delta k(\omega, \omega^\prime) = k_p(\omega+\omega^\prime) - k_s(\omega) - k_i(\omega^\prime) - k_{QPM}$. A constant factor $k_{QPM}$ takes into account additional quasi-phase matching.
The parametric gain $\Gamma$ is proportional to the pump field amplitude and the second-order nonlinear susceptibility and determines the interaction strength. The operators $\hat{a}^\dagger(\omega)$ and $ \hat{b}^\dagger(\omega^\prime)$ correspond to the signal and idler subsystems, respectively.

In the Heisenberg picture, the output operators can be found as $\hat{a}(\omega,z) = \hat{\mathcal{U}}^\dagger(z) \hat{a}(\omega)\hat{\mathcal{U}}(z)$, where the unitary spatial-evolution operator is given by $ \hat{\mathcal{U}}(z) = \mathcal{T}_z \exp \left( i \int_0^z dz^\prime \hat{G}(z^\prime)  \right) $. Here, the symbol $\mathcal{T}_z$ denotes the spatial-ordering: i.e., it reorders the operators in ascending order from right to left with respect to the parameter $z$ in the exponential expansion and thus provides an exponential that is $z$-ordered. For the propagator in the form of Eq.~\eqref{eq_generator_full}, the Heisenberg equation leads to a system of coupled integro-differential equations for the signal and idler operators, which have a solution in the form of a Bogoliubov transformation with transfer functions that, in turn, obey a system of coupled integro-differential equations \cite{Kopylov_2025,Christ_2013} (see also supplementary material (SM) sec.~\ref{Spatial_Heisenberg} and ~\ref{exact}). This system can be solved numerically, providing a rigorous solution to the problem. Using this solution, we build the photon-number operators for the signal and idler fields at the output of the crystal of length $L$ as $ \hat{n}_{s}(\omega) \equiv \hat{a}^\dagger(L,\omega)\hat{a}(L,\omega) $ and $ \hat{n}_{i}(\omega)  \equiv \hat{b}^\dagger(L,\omega)\hat{b}(L,\omega) $, respectively. The spectral distributions of the signal and idler fields are obtained by averaging the photon-number operators over the initial vacuum state 
\begin{equation}    
     n_{s}(\omega) \equiv \braket{ \hat{n}_{s}(\omega) } ~  \text{ and } ~ 
     n_{i}(\omega) \equiv \braket{ \hat{n}_{i}(\omega) }.  
     % = \int d\omega'|F_{a,b}(\omega,\omega')|^2,
\end{equation}
The total averaged numbers of generated photons for the signal $N_s$ and idler $N_i$ fields are defined as $N_{s,i} =\int d\omega \: n_{s,i}(\omega) $. Note that for type-II PDC $N_s = N_i \equiv N$. In what follows, we use the total averaged number of generated photons $N$ as a measure of the parametric gain $\Gamma$. In turn, the joint spectral intensity (JSI) is defined as 
\begin{equation}
    \mathrm{JSI}(\omega, \omega^\prime) \equiv \braket{ \hat{n}_{s}(\omega) \hat{n}_{i}(\omega^\prime)}  - \braket{\hat{n}_{s}(\omega)}\braket{ \hat{n}_{i}(\omega^\prime)} .
\end{equation}
The explicit expressions of the averaged number of photons and JSI in terms of Bogoliubov functions are given in the SM sec.~\ref{appendix PDC properties}.

\paragraph*{Phase-matching condition.}
The properties of type-II PDC are determined by two factors: the properties of the pump and the dispersion of the waveguide. In this work, we limit ourselves to dispersion up to the second order, assuming that the zero-order term of the wavevector mismatch $\Delta k_0 = k_p(\omega_0) - k_s(\omega_{0}/2)-k_i(\omega_{0}/2) -k_{QPM} \equiv 0$, where $\omega_0$ is the central frequency of the pump. This can be done via a proper choice of the quasi-phase matching $k_{QPM}$. Therefore, the wavevector mismatch under study has the form  
\begin{equation}
\label{eq_deltak_taylor_expansion}
    \Delta k\left(\omega_s,\omega_i\right) = \Delta k_1(\Omega_s, \Omega_i) + \Delta k_2(\Omega_s, \Omega_i),
\end{equation}
where $\Omega_{s,i}=\omega_{s,i}-\omega_{0}/2$ is the frequency detuning. 
The first-order phase-matching term can be written as 
\begin{equation}
    \Delta k_1(\Omega_s, \Omega_i) = \alpha_s \Omega_s + \alpha_i \Omega_i,
    \label{eq:Delta_k1}
\end{equation}
where the parameters $\alpha_{s,i}$ are defined by the group-velocities $v^g$ of interacting fields as $\alpha_{s,i} = 1/{v^g_p}-1/v^g_{s,i}$. In turn, the second-order term is defined by the group-velocity dispersion $\beta_{s,i,p} \equiv d^2k_{s,i,p}/d\Omega^2|_{\Omega=0}$ of the interacting fields as
\begin{multline}\label{deltak2}
     \Delta k_2(\Omega_s, \Omega_i) = \frac{1}{2}\Big(\beta_p(\Omega_s +\Omega_i)^2  -\beta_s\Omega_s^2-\beta_i\Omega_i^2 \Big).
\end{multline}
As a result, the dispersion is parametrized by five parameters: $\alpha_s$, $\alpha_i$, $\beta_p$, $\beta_s$, and $\beta_i$, based on which the characteristic times of the system can be introduced. Namely, for a waveguide of length $L$, the first-order characteristic time can be defined as the well-known  group delay between the signal and idler fields $\tau_1 = |\alpha_s - \alpha_i|L$. The second-order characteristic time does not have a strict definition. To introduce it, we use the curvature $\kappa$ \cite{GOLDMAN2005632} of the curve $\Delta k(\omega_s, \omega_i)=0$ at  the point $\Delta k(\omega_0/2,\omega_0/2)$, namely, $\tau_2 = |\kappa|$. The explicit expression for the curvature is given in the SM sec.~\ref{app_second_time}.

\paragraph*{Criteria for PDC degeneracy.}
The frequency-degenerate type-II PDC process constitutes a regime in which the signal and idler photons are generated at the same central frequency $\omega_0/2$, even if they have different spectral profiles. Usually, this situation corresponds to vanishing wavevector mismatch at the mentioned frequency, namely, $\Delta k (\omega_0/2,\omega_0/2) \equiv 0$.
Note that this condition is always fulfilled for wavevector mismatch given in the form of Eq.~\eqref{eq_deltak_taylor_expansion}.
However, as we will show below, in highly dispersive media, fulfilling this condition does not guarantee the degeneracy of the signal and idler spectra in the high-gain regime.
To quantify the degree of \textit{non-degeneracy},
we use the \textit{relative spectral distance}
\begin{equation}
   \mathrm{RSD}[n_s(\omega), n_i(\omega)]  = \dfrac{|\bar{\omega}_s-\bar{\omega}_i|}{\mathrm{max}(\sigma_s, \sigma_i)},
    \label{eq_rsd_definition}
\end{equation}
where the frequencies $\bar{\omega}_s$ and $\bar{\omega}_i$ are the characteristic central frequencies of the spectra  $\bar{\omega}_{s,i}=\frac{1}{N} \int d\omega \: \omega \: n_{s,i}(\omega)  $ and the values $\sigma_s$ and $\sigma_i$ characterize the widths of the signal and idler spectra in terms of the standard deviation $\sigma_{s,i} = \sqrt{ \frac{1}{N} \int d\omega \: (\omega-  \bar{\omega}_{s,i})^2 n_{s,i}(\omega) } $.
According to this definition, $\mathrm{RSD}=0$
corresponds to \textit{degenerate} PDC, while the higher the RSD is, the larger the spectral non-degeneracy becomes. Note that other definitions can be used to determine characteristic parameters in Eq.~\eqref{eq_rsd_definition}, e.g, for the spectral width, the full width at half maximum (FWHM) can be considered, and for the characteristic central frequencies, the frequencies corresponding to the spectrum's maximum.

Below, to check the spectral distinguishability,  we will also pay attention to the \textit{spectral overlap} of the signal and idler beams
\begin{equation}
  \Theta[n_s(\omega), n_i(\omega)] = \dfrac{\int d\omega n_s(\omega) n_i(\omega)}{\sqrt{\int d\omega n^2_s(\omega)} \sqrt{\int d\omega n^2_i(\omega)}}.
\end{equation}
Note that the spectral overlap does not generally indicate spectral degeneracy, as the overlap may be small for degenerate spectra when the widths differ significantly.

\paragraph*{A set of waveguides with various dispersion.}
\begin{figure}[!ht]
    \includegraphics[width=0.99\linewidth]{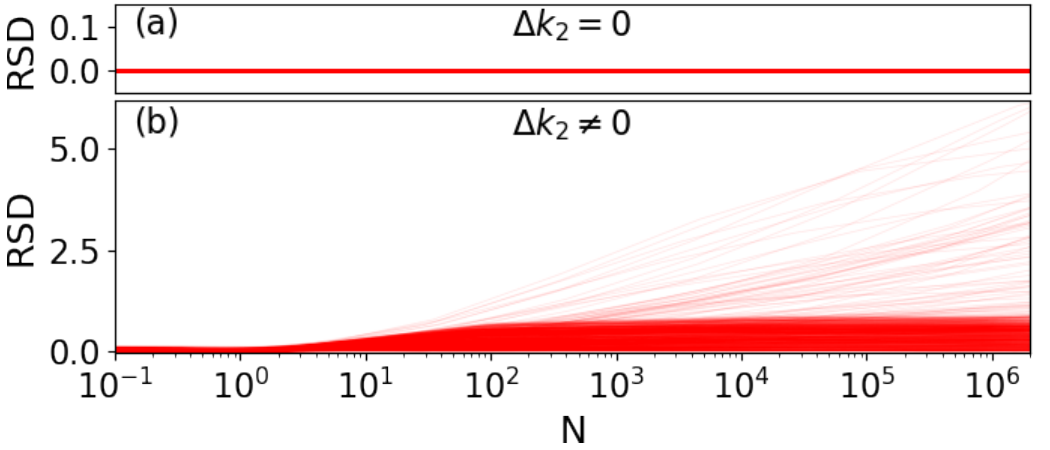} 
    \includegraphics[width=0.99\linewidth]{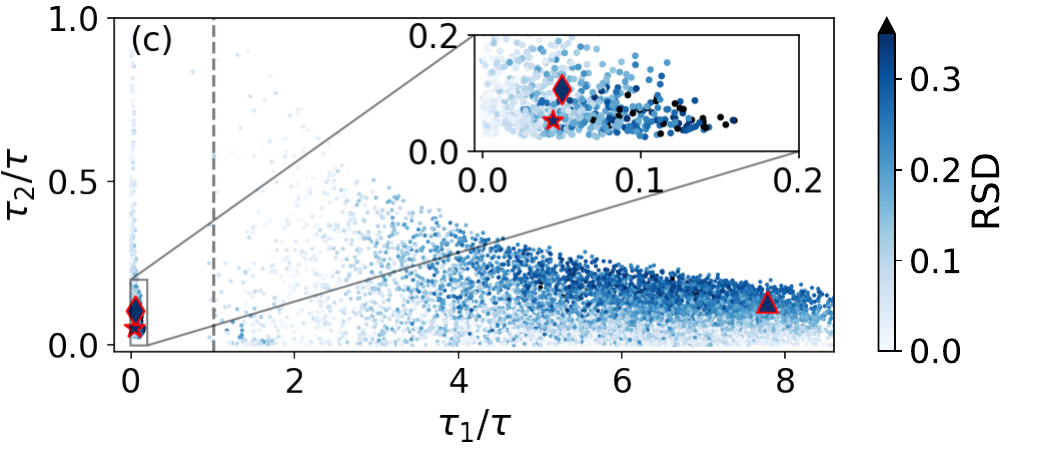} \\
    \includegraphics[width=\linewidth]{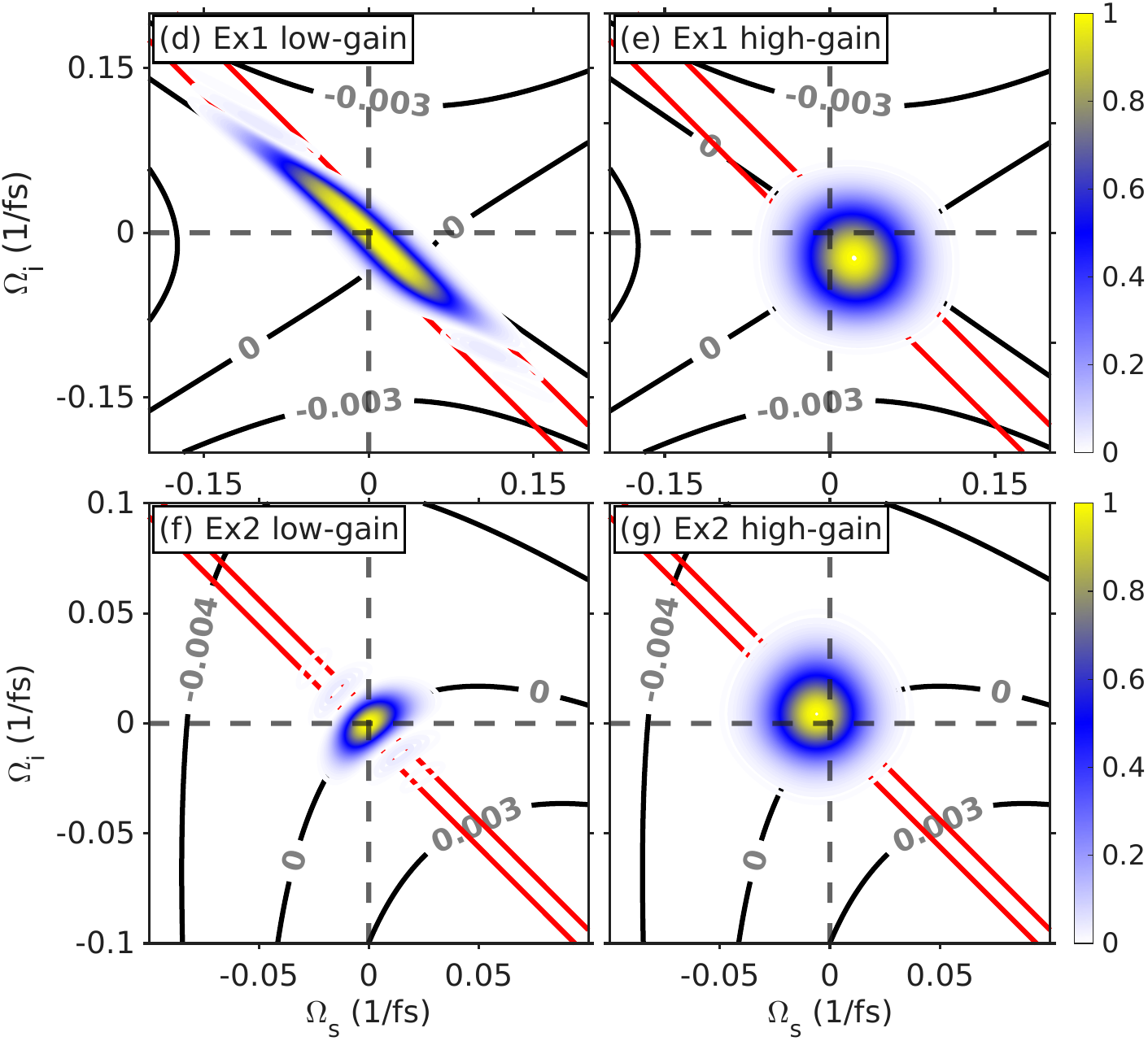} 
    \caption{(a,b) The dependence of the RSD on the total number of generated photons  (parametric gain) for a set of waveguides with randomly chosen dispersion in the case of (a) $\Delta k_2=0$ and (b) $\Delta k_2\neq0$.
    Each line presents a fixed waveguide. 
    (c) The distribution of the RSD (color bar) in the characteristic time space ($\tau_1/\tau$, $\tau_2/\tau$) for the fixed number of generated photons $N=10$.
    Each dot presents a fixed waveguide. The diamond and triangle correspond to examples Ex1 and Ex2 shown in (d, e) and (f, g), respectively. (d, f) The low-gain regime with ($N \approx10^{-5}$), (e, g) the high-gain regime with ($N\approx10^{7}$).  The star corresponds to the random waveguide in  Fig.~\ref{evol} of the SM sec.~\ref{evo_section}. 
    In Figs. (d-g) and below for similar figures, the red diagonal lines indicate the FWHM of the pump spectral function $S(\Omega_s+\Omega_i)$. The black solid lines represent the isolines of the wavevector mismatch $\Delta k$ [1/$\mu m$]. 
    }
    \label{fig_2_random}
\end{figure}

To demonstrate the effect of gain-induced non-degeneracy, we consider a set of waveguides of length $L=1$~cm, for which we randomly choose the dispersion parameters $\alpha_s$, $\alpha_i$, $\beta_p$, $\beta_s$, and $\beta_i$ using the sampling procedure given in the SM sec.~\ref{Protocol}. In addition, we fix the pump parameters as follows: $\lambda_p=0.775$~$\mu$m for the central wavelength and $\tau=80$~fs for the pulse duration. For each waveguide, we numerically solved the system of coupled integro-differential equations (see SM sec.~\ref{exact}) and calculated the $\mathrm{RSD}$; the results are shown in Fig.~\ref{fig_2_random}. Note that to uniquely determine the $\mathrm{RSD}$, we trim off the low-intensity tails of the photon-number distribution, when present; for details see SM sec.~\ref{Characteristic parameters}. Figs.~\ref{fig_2_random}(a)-(b) illustrate the behavior of the $\mathrm{RSD}$ as a function of the total number of generated photon pairs.
Considering waveguides with zero second-order dispersion terms, namely, $\beta_p=\beta_s=\beta_i = 0$ (which leads to  $\Delta k_2 = 0$ and  $\tau_2 = 0$), we get $\mathrm{RSD}\equiv 0$ for all photon-number values as demonstrated in
Fig.~\ref{fig_2_random}(a). In turn, Fig.~\ref{fig_2_random}(b) presents the RSD for non-zero second-order dispersion (non-zero $\tau_2$). Note that to obtain these results, we used an additional restriction on the waveguide samples: Only the dispersion parameters that provide the almost degenerate low-gain PDC ($\mathrm{RSD}(N \rightarrow 0)< 0.15$) with the high spectral overlap ($\Theta>0.96$) are taken into account.
Despite the extremely close spectra in the low-gain regime, we observe a significant increase in the $\mathrm{RSD}$ for many configurations as the parametric gain increases, indicating the presence of gain-induced non-degeneracy.
Fig.~\ref{fig_2_random}(c) shows the dependence of the $\mathrm{RSD}$ on the characteristic times normalized to the pump duration for a fixed number of generated photons $N=10$ (the full-size image is given in the SM sec.~\ref{Sampling_protocol}). Here, one can distinguish two main regions:  1)  $\tau_1/\tau <<1$ and 2) $\tau_1/\tau >1$. Note that the pure white areas appear due to the chosen sampling procedure: For them, the JSI does not satisfy at least one of the conditions in the low-gain regime: a) localization in the considered frequency range, b) small degeneracy, c) high spectral overlap; the bounds for the last two criteria are identical to those in Fig.~\ref{fig_2_random} (b). 

In the first region, the  $\mathrm{RSD}$ increases when  $\tau_1$ grows, in the second region, an opposite tendency is observed: the  $\mathrm{RSD}$ increases with increasing  $\tau_2$. The JSIs of two waveguides marked by the diamond and triangle that belong to the two described regions are presented in Figs.~\ref{fig_2_random} (d, e) and Figs.~\ref{fig_2_random} (f, g), respectively; their curvatures can be found in the SM sec.~\ref{app_second_time}. To understand such RSD behavior, one should examine the curvature of the phase-matching surface: With increasing gain, the maximum of the JSI distribution tends to shift toward the point with maximal curvature of the  $\Delta k=0$ isoline (PM0 curve), provided that this point belongs to the region where the pump spectral function is still not vanishing. For this curve, $\tau_1$ determines the slope at the point of origin, while $\tau_2$ describes the curvature at the same point. Therefore, in the first region, in the point of origin, the PM0 curve is oriented almost along the pump diagonal ($\Omega_s+\Omega_i=0$) and has a large overlap with the pump function; in the second region, this curve is oriented at some non-zero angle to the pump diagonal and, thus, a smaller part of PM0 curve overlaps with the pump spectral function. As a result, to maximally shift the point with the maximal curvature of the PM0 curve from the point of origin and simultaneously maintain its greater overlap with the pump spectral function (conditions required for non-zero RSD), $\tau_1$ should be increased in the first region, and $\tau_2$ in the second. 
 
\paragraph*{A particular example.}
\begin{figure}[h]
    \includegraphics[width=\linewidth]{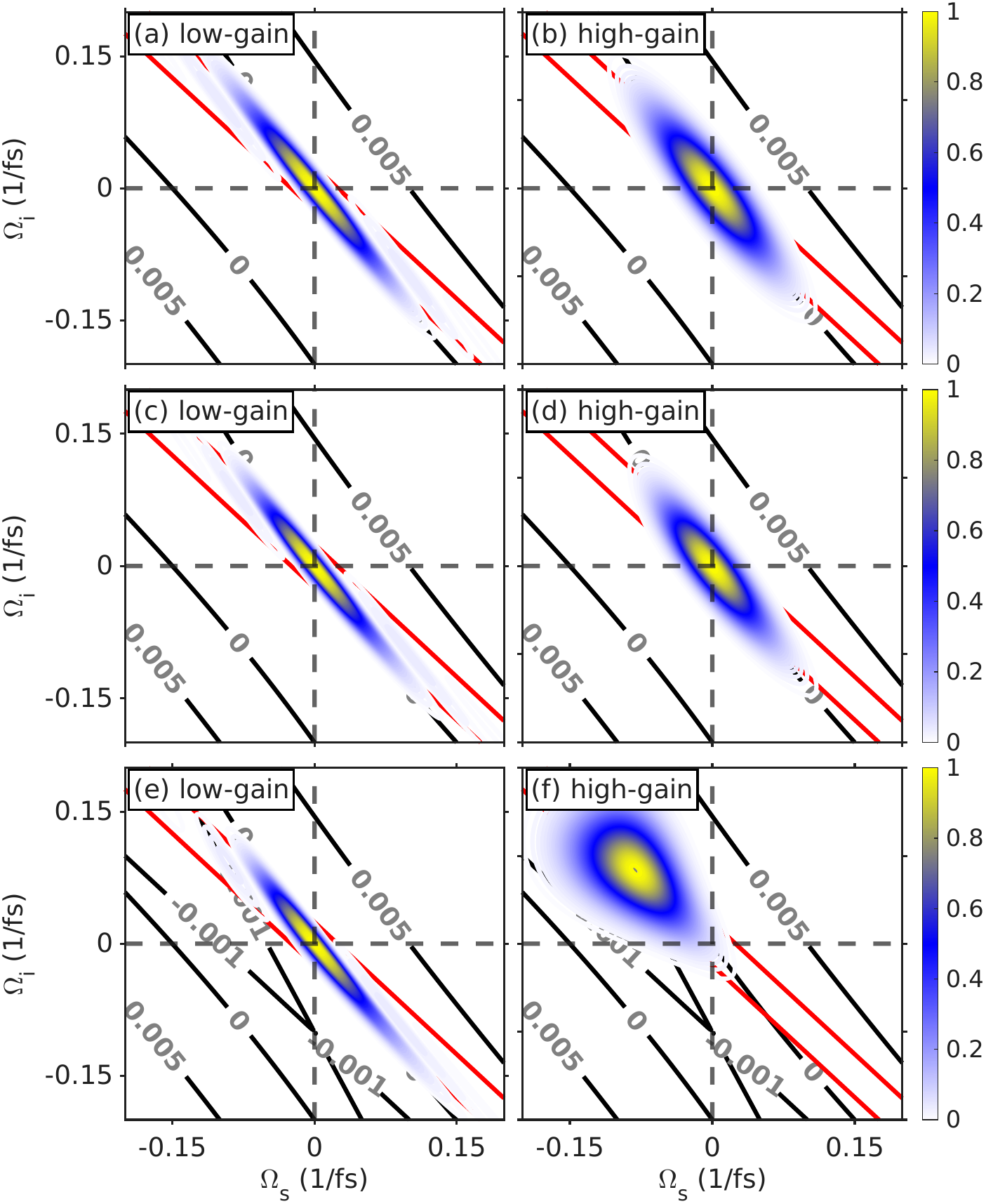}
    \caption{Normalized JSI for (a,b) WG0, (c,d) WG1, and (e,f) WG2 in the case of (a,c,e) low-gain and (b,d,f) high-gain regimes with
    the total number of generated photons  $N\approx10^{-5}$ and $N\approx10^5$, respectively. 
   }
\label{fig_JSI_rigor}
\end{figure}

To illustrate gain-induced non-degeneracy in more detail and highlight the importance of high-order dispersion terms in high-gain PDC, we consider three particular waveguides with the same linear phase-matching term $\Delta k_1$ but different second-order terms: WG0 ($\Delta k_2$=0), WG1 (small dispersion), and WG2 (strong dispersion). The JSIs for the three waveguides are presented in Fig.~\ref{fig_JSI_rigor}, and the phase-matching parameters are given in Table~\ref{params}.

\begin{table}[h]
    \centering
        \resizebox{0.48\textwidth}{!}{
\begin{tabular}{c c c c c c c c}
           \hline  \hline  \noalign{\vskip 0.3ex}
           &$\alpha_s$ [$\frac{\textrm{fs}}{\textrm{mm}}$] &~$\alpha_i$ [$\frac{\textrm{fs}}{\textrm{mm}}]$&~$\beta_p$ [$\frac{\textrm{fs}\textsuperscript{2}}{\textrm{mm}}$]&~$\beta_s$ [$\frac{\textrm{fs}\textsuperscript{2}}{\textrm{mm}}$]&~$\beta_i$ [$\frac{\textrm{fs}\textsuperscript{2}}{\textrm{mm}}$]&~$\tau_1$ [\textrm{fs}]&~$\tau_2$ [\textrm{fs}]\\ [0.4ex]
           \hline    
    WG0 &   30&  20 &  0 &   0 & 0  &100& 0
    \\
   
    WG1 &   30&  20 &  30 &   10 & 10& 100& 0.21 \\
      
    WG2 &   30&  20 &  300 &   -100 & 100 & 100 & 0.42\\
    \hline \hline
    \end{tabular}
    }
    \caption{Phase-matching parameters used for Fig.~\ref{fig_JSI_rigor} together with  corresponding characteristic times. }
    \label{params}
\end{table}

For this particular example, we selected the parameters such that all three waveguides have similar JSIs in the low-gain regime (their spectral overlap is $ \Theta$ $ > 0.9859$); see Figs.~\ref{fig_JSI_rigor}(a,c,e). To understand why three very different waveguides behave similarly in the low-gain regime, one can estimate the first- and second-order phase-matching terms according to Eqs.~(\ref{eq:Delta_k1}) and (\ref{deltak2}), respectively, substituting the pump characteristic time (pulse duration) as $\Omega_s = -\Omega_i = 1/\tau$, namely, $\Delta k_1^{\mathrm{est}} =(\alpha_s-\alpha_i)/ \tau= 0.125 \ \mathrm{mm}^{-1}$ and  $\Delta k_2^{\mathrm{est}} = -(\beta_s + \beta_i)/ (2\tau^2)=-0.0015625 \ \mathrm{mm}^{-1}$ for  WG1  and $\Delta k_2^{\mathrm{est}} = 0 \ \mathrm{mm}^{-1}$ for WG2, which clearly demonstrates that for all waveguides the first-order phase-matching term dominates. This may create the wrong intuition that the second-order phase-matching term can be neglected for all three waveguides, which is, however, incorrect. Figs.~\ref{fig_JSI_rigor}(b,d,f) show that the waveguides have dramatically different JSIs in the high-gain regime, namely, the JSI of WG2 is shifted in the frequency space. This shift indicates a gain-induced non-degeneracy as shown in Fig.~\ref{fig_JSI_rigor}(f), which is not present for WG0 and WG1 as demonstrated in  Figs.~\ref{fig_JSI_rigor}(b,d). Such different behavior is explained by a strong difference in the phase-matching curvature profile of the waveguides (see  SM sec.~\ref{app_second_time}) which is caused by significantly different second-order dispersion parameters: Each individual second-order dispersion term of WG1 is much smaller than the linear one  $\beta_{s,i,p}/ \tau^2 << \Delta k_1^{\mathrm{est}}$, however, those of  WG2 are already comparable with $\Delta k_1^{\mathrm{est}}$. Note that the phase-matching curvature and the characteristic time $\tau_2$ also affect the rate of the shift with increasing the number of photons: the larger the curvature is, the faster the shift occurs.
 
In addition, since the gain-induced non-degeneracy depends on both the phase-matching characteristics (curvature and slope) and the pump duration, there may be cases where the RSD does not occur in the high-gain regime, although the second-order dispersion terms are present. In this case, the PM0 curve is oriented almost perpendicular to the pump diagonal, and the spectral bandwidth of the pump prevents the gain-induced frequency shift from occurring, see SM. sec.~\ref{without_shift} for details.

\begin{figure}[h]

    \includegraphics[width=0.49\linewidth]{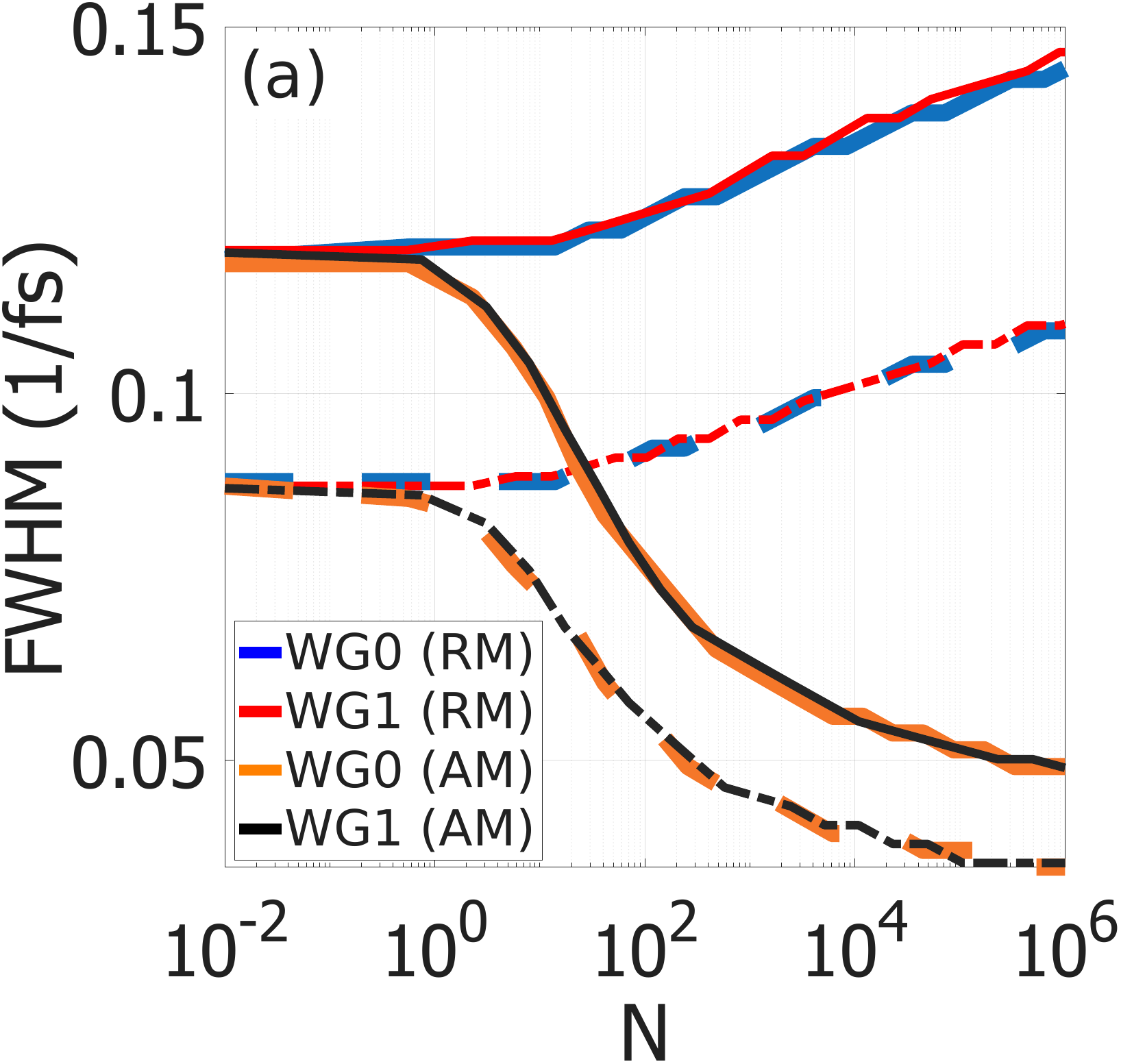}
    \includegraphics[width=0.49\linewidth]{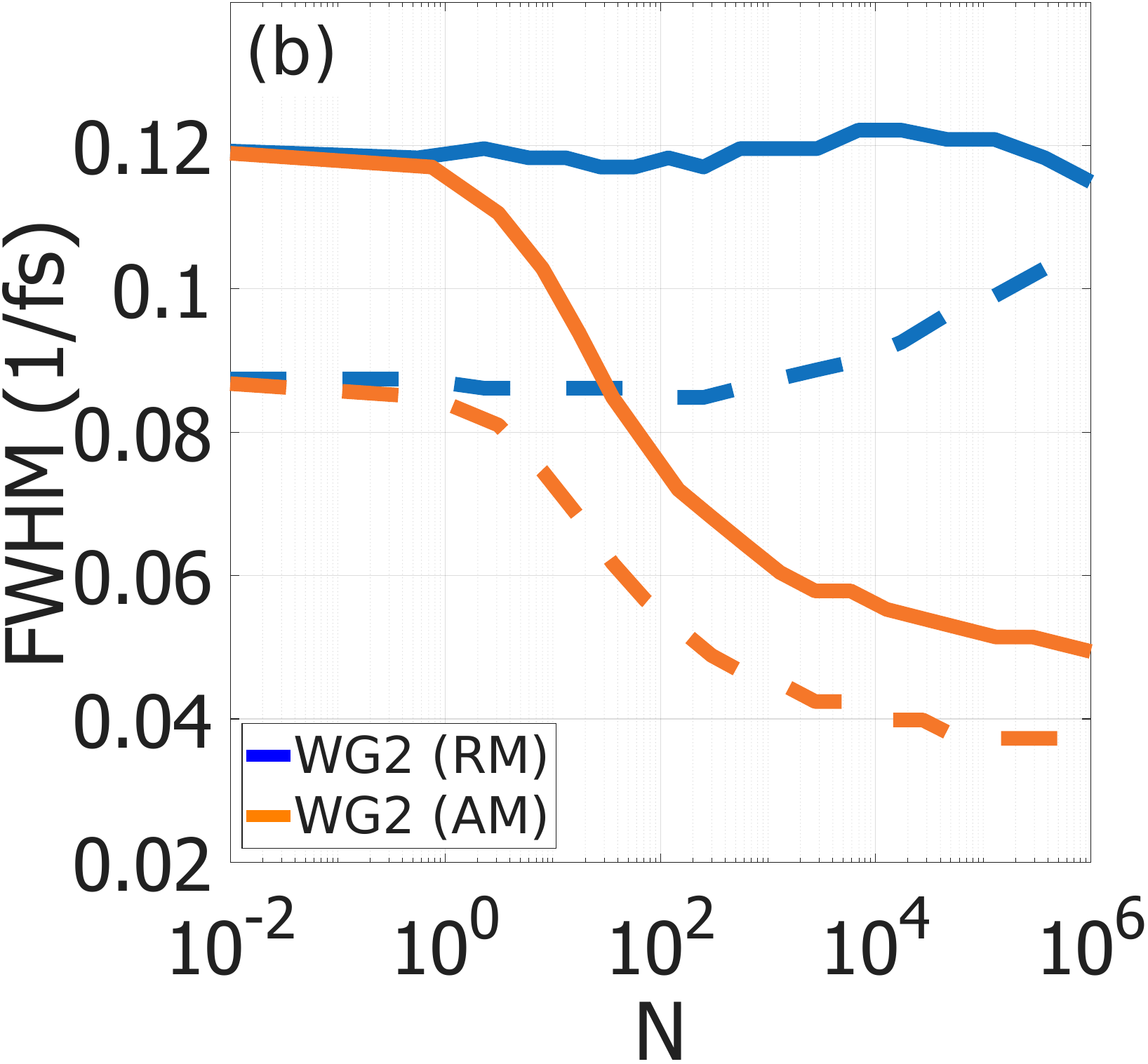}\\
    \caption{FWHM of the idler (solid) and signal (dashed) spectra as a function of the number of generated photons for (a) WG0 and WG1 and (b) WG2. RM corresponds to the rigorous model and AM to the spatially-averaged model. }
    \label{fig_broadening_spectrum}
\end{figure}

The frequency shift is not the only unexpected property of light generated in systems with strong dispersion. Fig.~\ref{fig_broadening_spectrum} demonstrates the behavior of the FWHM of the spectral distribution of the signal and idler beams with respect to the number of generated photons. For WG0 and WG1, where the second-order dispersion is small (Fig.~\ref{fig_broadening_spectrum}(a)), the well-known broadening of the spectrum is observed \cite{Sharapova_2020}. However, in the presence of strong second-order dispersion for WG2, a complicated modification of the spectra takes place as the gain increases, see SM sec.~\ref{evo_section}. Such modification results in a generally non-monotonic behavior of spectral width with respect to the number of generated photons, as shown in  Fig.~\ref{fig_broadening_spectrum}(b). For comparison, in Figs.~\ref{fig_broadening_spectrum} we also present the FWHM obtained using the averaged model ~\cite{Wasilewski_2006,Sharapova_2015} (see also SM sec.~\ref{AM}) that exhibits a completely different behavior (narrowing of signal and idler spectra) compared to the rigorous solution in the high-gain regime.

\begin{figure}[h]
    \includegraphics[width=1\linewidth]{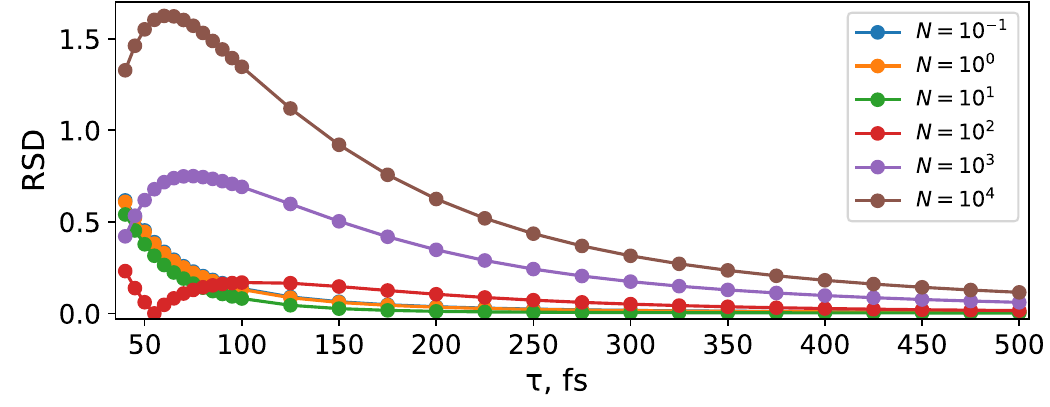}
    \caption{Relative spectral distance for WG2 as a function of the pump pulse duration. Different colors correspond to different numbers of generated photons N. Note that the blue and orange lines almost coincide.
    }
    \label{fig_pulse_duration}
\end{figure}

\paragraph*{Influence of pulse duration.} 
The next parameter that strongly affects the RSD is the duration of the pump pulse. Taking into account the estimates for the first- and second-order phase-matching terms using the pulse duration discussed above, one can recognize that the longer the pulse duration is, the stronger the second-order phase-matching term is suppressed compared to the first-order counterpart, which, in turn, results in the suppression of the gain-induced non-degeneracy with increasing the pump duration; see Fig.~\ref{fig_pulse_duration}. One can notice that the effect almost disappears for a $\tau=500$~fs pump.  Moreover, if the pump pulse duration exceeds 200~fs, the RSD monotonically increases with increasing the gain. However, for short pump pulses, a complicated dynamics takes place that leads to a non-monotonic behavior of RSD with the gain and requires further investigation.

\paragraph*{PPKTP waveguide.}
As demonstrated in Fig.~\ref{fig_2_random}, gain-induced non-degeneracy is not a rare effect. However, to the best of our knowledge, this effect has never been described either theoretically or experimentally, except for Ref.~\cite{Triginer2020}, where similar effects were explained by considering self-phase modulation. The reason is that experimentally obtaining this effect requires simultaneously highly dispersive media, short femtosecond pump pulses, and high gain. Conversely, to catch this effect theoretically, one needs to properly treat high-order dispersion terms in the high-gain regime within a rigorous numerical solution of the problem. Note that for the case of a continuous-wave pump, when the system of integro-differential equations is replaced by a system of differential equations that has an analytical solution, the gain-induced non-degeneracy is not present due to the delta-correlations between the signal and idler photons, see Fig.~\ref{fig_pulse_duration}. 

\begin{figure}[h]
\includegraphics[width=\linewidth]{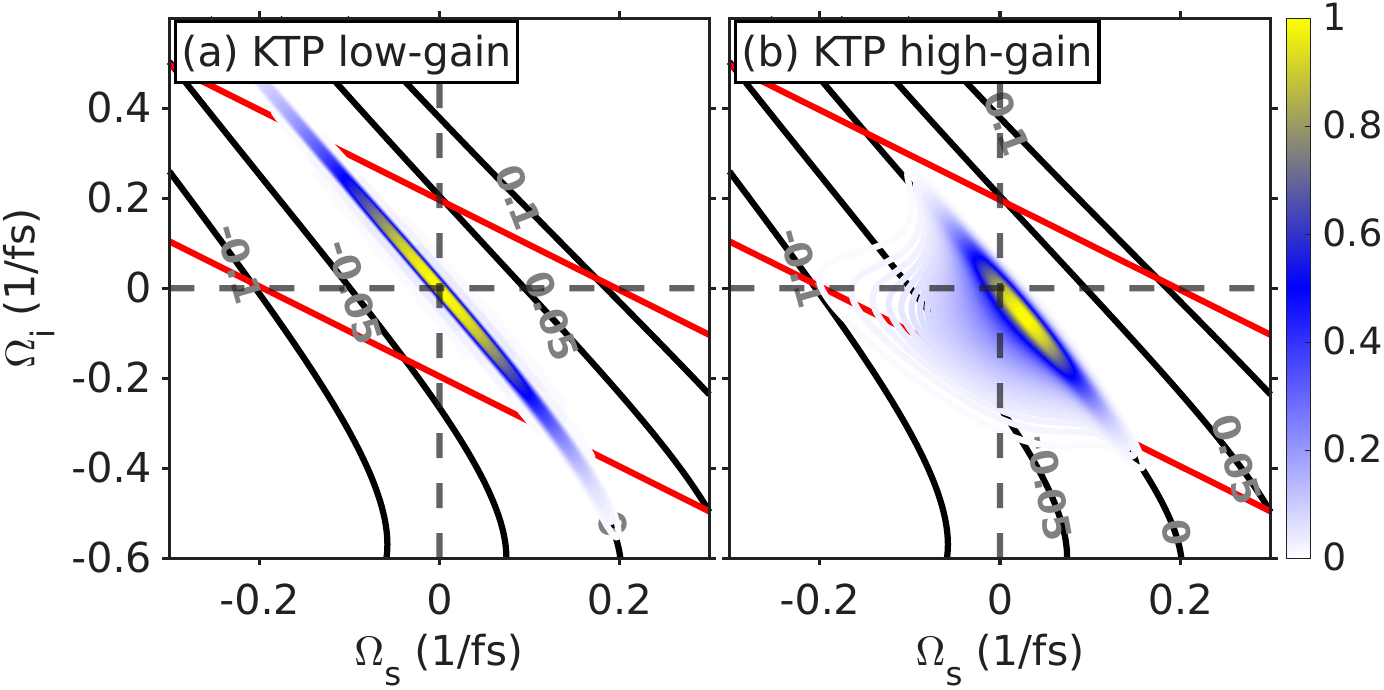}
    \caption{Normalized JSI for the PPKTP waveguide of length 1 mm for (a) low-gain ($N=6 \times10^{-5}$) and (b) high-gain ($N=1.6\times10^{6}$) regimes, the pump pulse duration is 10~fs. 
    }
    \label{KTP_exact}
\end{figure}

Finding waveguides with strong dispersion can be challenging; therefore, to observe the gain-induced non-degeneracy effect, a not-so-large dispersion can be compensated for by considering shorter pump pulses. To demonstrate this, we consider a type-II PPKTP waveguide with a poling period of $10.8 \mu m$ to close the $\gamma \rightarrow (\beta, \gamma) $ phase-matching synchronism. The length of the waveguide is $L=1$ mm, the expressions for the refractive indices are chosen to be close to the bulk material \cite{Kato:02}, the pulse duration is $\tau=10$~fs, which is shorter than in our previous calculations. The resulting JSIs in the low- and high-gain regimes are presented in Fig.~\ref{KTP_exact}, demonstrating the gain-induced non-degeneracy of the order of $\mathrm{RSD}=0.424$ for $N \approx10^{6}$ generated photons. The same figures calculated using the spatially-averaged model are presented in the SM sec.~\ref{PPKTP}.

\paragraph*{Summary.}
We have theoretically demonstrated the novel effect of gain-induced non-degeneracy for high-gain type-II PDC that can be observed in highly dispersive media with the use of ultra-short pump pulses. This special regime of generation can only be correctly described using the rigorous model (the widely used spatially-averaged model fails to reveal this regime) and taking into account the second-order dispersion terms (the effect cannot be predicted from theories that only consider linear dispersion terms related to group velocities). We have demonstrated that in this PDC regime, a complicated modification of the signal and idler spectra occurs, while the gain-induced non-degeneracy can be observed in a variety of waveguides. Recent progress in the fabrication of waveguides and fibers with modeled dispersion, such as LNOI waveguides \cite{Poberaj2012}, and microstructured fibers \cite{RevModPhys.89.045003}, opens up new possibilities for experimentally realizing this effect. Spectral manipulation through gain can find many applications in spectral engineering,  continuous-variable quantum computing, and quantum information protocols.  

\begin{acknowledgments}
 We acknowledge financial support of the Deutsche Forschungsgemeinschaft (DFG) via the TRR 142/3 (Project No. 231447078, Subproject No. C10).
\end{acknowledgments}

\bibliography{refs}

\newpage
~
\newpage

\onecolumngrid

\section*{Supplementary Material}

\subsection{Theoretical approach}\label{Theoretical approach}

\subsubsection{Spatial Heisenberg equations for type-II PDC}
\label{Spatial_Heisenberg}

The detailed derivation of the spatial Heisenberg equations for monochromatic creation and annihilation operators can be found in e.g.~\cite{Yang2008}; therefore here we only shortly introduce the main results. Having the slow-varying annihilation operators  for the signal $\hat{a}(\omega)$ and  idler $\hat{b}(\omega)$ fields (with commutation relations $[\hat{a}(\omega), \hat{a}^\dagger(\omega^\prime)]=\delta(\omega-\omega^\prime)$, $[\hat{b}(\omega), \hat{b}^\dagger(\omega^\prime)]=\delta(\omega-\omega^\prime)$, $[\hat{a}(\omega), \hat{b}^\dagger(\omega^\prime)]=0$), the spatial Heisenberg equations for the type-II PDC have the form \cite{Kopylov_2025,Christ_2013}
\begin{equation}\label{CIDE1}
\begin{split}
    \frac{d\hat{a}(\omega,z)}{dz}&= i \: \Gamma \int d\omega^\prime J(\omega, \omega^\prime, z)  b^\dagger(\omega^\prime,z), 
    \\
    \frac{d\hat{b}(\omega,z)}{dz}&= i \: \Gamma \int d\omega^\prime J(\omega^\prime, \omega, z)  a^\dagger(\omega^\prime,z),
    \end{split}
\end{equation}
where $J(\omega, \omega^\prime, z) =  S(\omega + \omega^\prime) e^{i\Delta k(\omega, \omega^\prime) z}$ is the $z$-dependent coupling function.  Here, $S(\omega+\omega^\prime)$ is the pump spectrum, $\Delta k(\omega, \omega^\prime)=k_p(\omega+\omega^\prime) - k_s(\omega) - k_i(\omega^\prime) - k_{QPM}$ is the wavevector mismatch, where $k_{s,i,p}(\omega) = \frac{n_{s,i,p}(\omega)\omega}{c}$ are the frequency-dependent wavevectors of the signal (s), idler (i), and pump (p) fields. A constant factor $k_{QPM}$ takes into account the additional quasi-phase matching. The parameter $\Gamma$ determines the interaction strength and is proportional to the pump field amplitude and the second-order nonlinear susceptibility.

The solution to the Heisenberg equations from $0$ to $z$ can be written as  $\hat{a}(\omega,z) = \hat{\mathcal{U}}^\dagger(z) \hat{a}(\omega)\hat{\mathcal{U}}(z)$, where $\hat{a}(\omega)\equiv\hat{a}(\omega,0)$ is the vacuum operator, and the spatial evolution operator $\hat{\mathcal{U}}(z)$ is a unitary operator of the form $ \hat{\mathcal{U}}(z) = \mathcal{T}_z \exp \left( i \int_0^z dz^\prime \hat{G}(z^\prime)  \right) $. The symbol $\mathcal{T}_z$ denotes the requirement of the exponential to be $z$-ordered (causality), i.e., the operators in the exponential expansion have an increasing order from right to left with respect to the parameter $z$. According to Eqs.~\eqref{CIDE1}, the spatial propagator has the form
\begin{equation}
    \hat{G}(z) = \Gamma \iint d\omega d\omega^\prime J(\omega,\omega^\prime,z) \hat{a}^\dagger(\omega) \hat{b}^\dagger(\omega^\prime)  + h.c. 
\end{equation}

\subsubsection{Rigorous model}
\label{exact}

Eqs~\eqref{CIDE1} are linear with respect to the annihilation and creation operators; therefore, their solution can be written in the form of the Bogoliubov transformation
\begin{equation}
        \hat{a}(\omega,z) = \int d\omega^\prime \left( E_a(\omega, \omega^\prime, z) \hat{a}(\omega^\prime)  +  F_a(\omega, \omega^\prime, z) \hat{b}^\dagger(\omega^\prime)  \right),
        \label{eq_bogoliubov_full_app1}
 \end{equation}
\begin{equation}
        \hat{b}(\omega,z) = \int d\omega^\prime \left( E_b(\omega, \omega^\prime, z) \hat{b}(\omega^\prime)  +  F_b(\omega, \omega^\prime, z) \hat{a}^\dagger(\omega^\prime)  \right),
        \label{eq_bogoliubov_full_app2}
\end{equation}
where $\hat{a}(\omega^\prime)\equiv\hat{a}(\omega^\prime,0)$ and $\hat{b}(\omega^\prime)\equiv\hat{b}(\omega^\prime,0)$ are the vacuum operators, while the Bogoliubov functions satisfy the following integro-differential equations
\begin{align}
    \dfrac{d E_{a}(\omega, \omega^\prime, z)  }{dz}  & =  i\Gamma \int d\omega^{\prime\prime} J(\omega, \omega^{\prime\prime}, z) F^\ast_b(\omega^{\prime\prime}, \omega^\prime, z),   
    \label{ap_eq_slowly_eq1}
    \\    
    \dfrac{d F_{b}(\omega, \omega^\prime, z)  }{dz} & =  i\Gamma \int d\omega^{\prime\prime} J(\omega^{\prime\prime}, \omega, z) E^\ast_a(\omega^{\prime\prime}, \omega^\prime, z),
    \label{ap_eq_slowly_eq2}
    \\    
    \dfrac{d F_{a}(\omega, \omega^\prime, z)  }{dz} & =  i\Gamma \int d\omega^{\prime\prime} J(\omega, \omega^{\prime\prime}, z) E^\ast_b(\omega^{\prime\prime}, \omega^\prime, z), 
    \label{ap_eq_slowly_eq3}
    \\
    \dfrac{d E_{b}(\omega, \omega^\prime, z)  }{dz} & =  i\Gamma \int d\omega^{\prime\prime} J(\omega^{\prime\prime}, \omega, z) F^\ast_a(\omega^{\prime\prime}, \omega^\prime, z),   
    \label{ap_eq_slowly_eq4}
\end{align}
with the initial conditions $E_{a}(\omega, \omega^\prime, 0)=E_{b}(\omega, \omega^\prime, 0)=\delta(\omega-\omega^\prime)$ and $F_{a}(\omega, \omega^\prime, 0)=F_{b}(\omega, \omega^\prime, 0)=0$.

The analytical solution of the presented equations does not exist: The main difficulty arises due to the explicit dependence of the coupling matrix $J(\omega, \omega^{\prime}, z)$ on $z$.
The numerical solution of Eqs.~\eqref{ap_eq_slowly_eq1}-\eqref{ap_eq_slowly_eq4} can be performed in a discrete frequency space, which results 
in a system of linear differential equations that can be solved numerically.
The main advantage of the direct numerical solution is that it obeys causality, i.e., the spatial-ordering effect is fully taken into account.

\subsubsection{PDC properties}\label{appendix PDC properties}
All required characteristics of PDC can be written in terms of the output Bogoliubov functions. For example, the averaged numbers of generated photons for the signal and idler beams can be written as \cite{Sharapova_2020}
 \begin{align}
   \braket{\hat{n}_{s}(\omega_s)}=\int d\omega'|F_a(\omega_s,\omega')|^2 ~\text{and}~  \braket{\hat{n}_{i}(\omega_i)}=\int d\omega'|F_b(\omega_i,\omega')|^2.
\end{align}
The  Joint Spectral Intensity (JSI) is given by 
\begin{align}
   JSI(\omega_s, \omega_i)= \int d\omega' d\omega'' F_a^\ast(\omega_s, \omega', z)E_b^\ast( \omega_i,\omega', z)E_a(\omega_s,\omega'', z)F_b(\omega_i, \omega'', z).
\end{align}

\subsection{Phase-matching curvature}\label{app_second_time}

\begin{figure}[h]
   \includegraphics[width=0.8\linewidth]{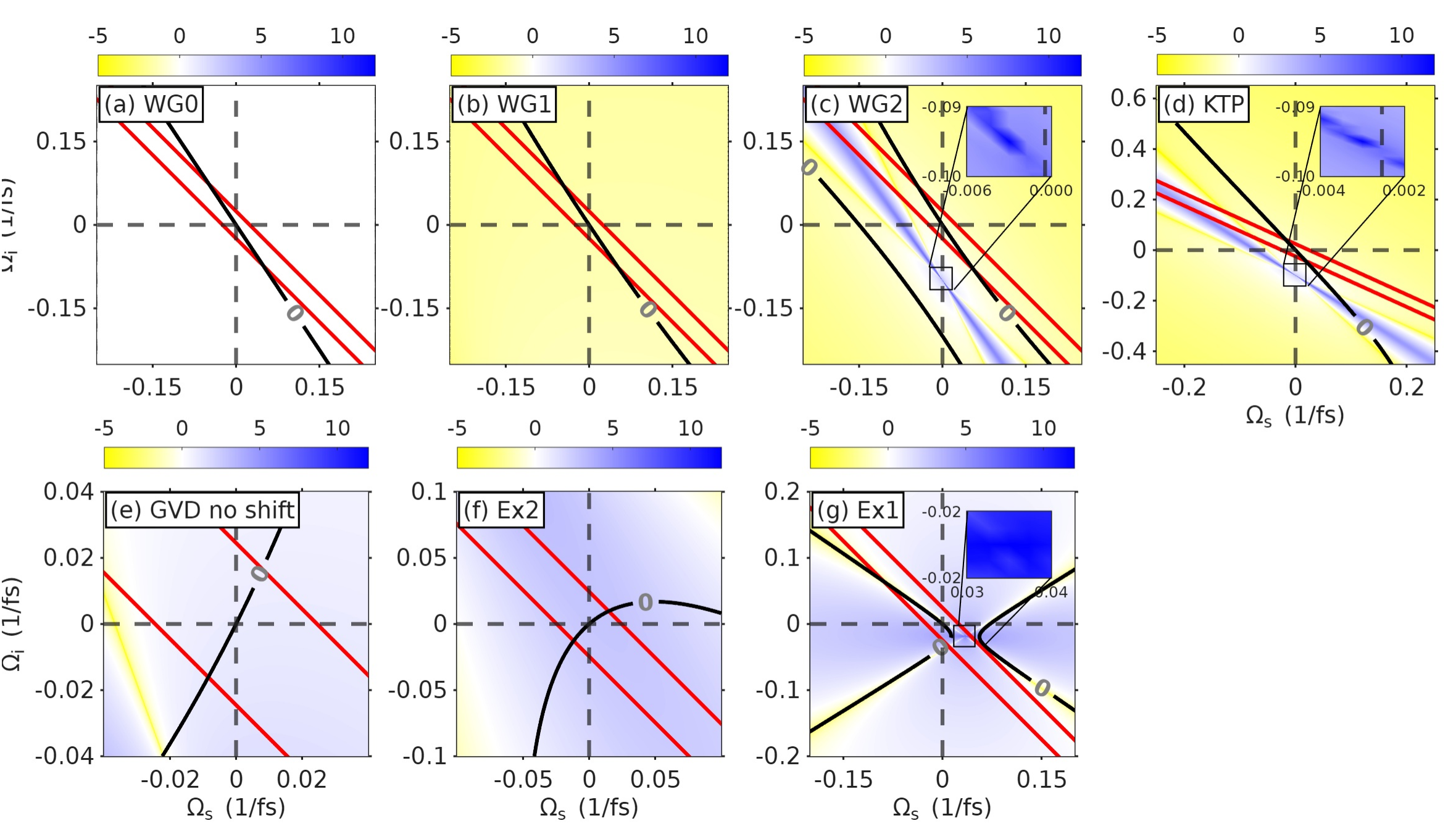}
    \caption{Curvature $\kappa$ [fs] for different waveguides presented in this work in logarithmic scale (excluding (a)).}
    \label{curv}
\end{figure}

For the implicitly defined curve $F(x,y) = 0$, the curvature  $\kappa$ is defined as \cite{GOLDMAN2005632}
\begin{equation}
    \kappa = \dfrac{F_{y}^2 F_{xx} - 2 F_x F_y F_{xy}+F_x^2F_{yy} }{ \left( F_x^{2}+F_y^{2}\right)^{3/2}}.
\end{equation}
For the $\Delta k(\Omega_s, \Omega_i)=0$,
the derivatives are: $F_x = F_{\Omega_s}=\alpha_s + (\beta_p-\beta_s) \Omega_s + \beta_p \Omega_i $, $F_y = F_{\Omega_i} = \alpha_i + (\beta_p-\beta_i) \Omega_i + \beta_p \Omega_s$, $F_{xx} = F_{\Omega_s\Omega_s} = \beta_p-\beta_s$, $F_{yy} = F_{\Omega_i\Omega_i} = \beta_p-\beta_i$ and $F_{xy} = F_{\Omega_s\Omega_i} = \beta_p$.
Then, at the point $\Omega_s=\Omega_i=0$ the curvature reads
\begin{equation}
    \kappa|_{\Omega_s=\Omega_i=0} = \dfrac{\alpha_i^2(\beta_p-\beta_s) - 2 \alpha_s \alpha_i \beta_p+\alpha_s^2(\beta_p-\beta_i) }{ \left( \alpha_s^2+\alpha_i^2\right)^{3/2}}.
\end{equation}
The curvature of waveguides used in this work is presented in Fig.~\ref{curv}.

\subsection{Figure 2: Details}\label{Sampling_protocol}
\subsubsection{Protocol for the Figure 2}
\label{Protocol}

For the Fig.~\ref{fig_2_random}, we used the following sampling procedure:
\begin{enumerate} 
    \item We fix the pump pulse duration to $80$~fs and the crystal length of $1$ cm.
    \item The parameters $\alpha_s$, $\alpha_i$ were taken independently from the uniform distribution within the range $[-35, 35]$~fs/mm.  
     \item We fix the frequency grid for the signal and idler components with the following parameters: the step size is $\delta \omega = 2\pi \times 0.36$ THz, the number of points for each signal and idler subsystem is equal to $M=255$. 
    The parameters $\beta_p$, $\beta_s$, and $\beta_i$ were taken independently from the uniform distribution within the range $[-350, 350]$~fs$^2$/mm.
\end{enumerate}

Since we have already fixed the frequency grid, for the correct numerical computation, we impose two restrictions on the generated PDC light: first, the spectrum of the generated PDC light should be localized; second, it must be resolved within the studied frequency interval. To choose suitable configurations, we used the following steps: 

\begin{enumerate}
    \item For the parameters generated at step (3), we calculate the JSI in the low-gain regime. This computation does not require the solution of differential equations, and, therefore, is very fast and cheap. 
    Considering the JSI in the low-gain regime, we used the following criteria to take the `good' samples from step (3):
        \begin{enumerate}
            \item The criterion for taking the PDC  spectra localized inside the studied grid
             $$  4 \times \delta \omega  < \Delta \omega_{s,i} < 0.1 \times M \times \delta \omega , $$ where  $\Delta \omega_s$ and $\Delta \omega_i$ are the spectral widths (FWHM) of the signal and idler beams.
            \item The criterion for taking the PDC spectra with resolved signal-idler correlations:
            $$ 2.5 \times \delta \omega < \sigma[C_{s,i}(\Omega_{s,i})] < 0.3 \times  M \times \delta \omega.  $$ Here, the $C_{s,i}(\Omega_{s,i})$ are the normalized cross-sections of the JSI : $C_s(\Omega_s) \propto JSI(\Omega_s, \Omega_i=0)$, $C_i(\Omega_i) \propto JSI(\Omega_s=0, \Omega_i)$.
            The standard deviation $\sigma[\cdot]$ characterizes the widths of these cross-sections.
         \end{enumerate}
\end{enumerate}

In Figure~\ref{fig_2_random}(a), we present the simulation results of PDC for $400$ random dispersion parameters holding  $\beta_{p,i,s}\equiv0$. 
Note that as long as $\Delta k_0 = 0$, the spectra are always centered around the central PDC frequency $\omega_0$. 

For a nonzero $\beta$, the condition $\Delta k_0 = 0$ does not guarantee that the PDC spectra are centered around $\omega_0$; they can have strongly asymmetric profiles. 
In order to study the gain-induced non-degeneracy in this case, we introduce two additional criteria for the PDC in the low-gain regime:
        \begin{enumerate}
            \item The first criterion is
                $$ \Theta > 0.96, $$ providing high overlap between the signal $n_s(\omega)$ and idler $n_i(\omega)$ spectra.  
                
            \item The second criterion is 
                $$ \mathrm{RSD} < 0.15. $$
        \end{enumerate}
        
Both these criteria significantly restrict the number of samples; however, they allow us to focus on the cases for which the signal and idler spectra have similar profiles in the low-gain regime. The results for $400$ samples are given in Figures~\ref{fig_2_random}(b).

\subsubsection{Characteristic parameters of the spectra}
\label{Characteristic parameters}

In order to uniquely define the RSD, instead of using the initial photon-number spectra $n_{s,i}(\omega_j)$, we use a modified function $ n^m_{s,i}(\omega_j)$
\begin{align}
    n^m_{s,i}(\omega_j) & = n_{s,i}(\omega_j), ~ ~  \text{if} ~ ~  n_{s,i}(\omega_j)>0.005\times \mathrm{max}[n_{s,i}(\omega_i)], \\
    n^m_{s,i}(\omega_j) & = 0 , ~ ~  \text{otherwise.}
\end{align}
As a characteristic central frequency, we use the averaged frequency which is defined as $\bar{\omega}_{s,i} = \frac{1}{N^m} \sum_j \omega_j n^m_{s,i}(\omega_j)$, where
$N_m = \sum_j n^m_{s,i}(\omega_j)$.
As a characteristic spectral width, we use the standard deviation which is given by  $\sigma^m_{s,i} = \sqrt{ \frac{1}{N^m} \sum_j \: (\omega_j-  \bar{\omega}_{s,i})^2 n^m_{s,i}(\omega_j) } $.

The usage of  $n^m_{s,i}(\omega_j)$ is helpful when the standard deviation of the initial photon-number spectra $n_{s,i}(\omega_j)$ depends on the choice of the size of numerical grid (e.g. when in the limit $\omega\rightarrow \infty$ the intensity distribution  $n_{s,i}(\omega)$ has some constant tails). 

\subsubsection{Figure 2}
\label{Fig2c_full}

\begin{figure}[h]
    \includegraphics[width=0.41\linewidth]{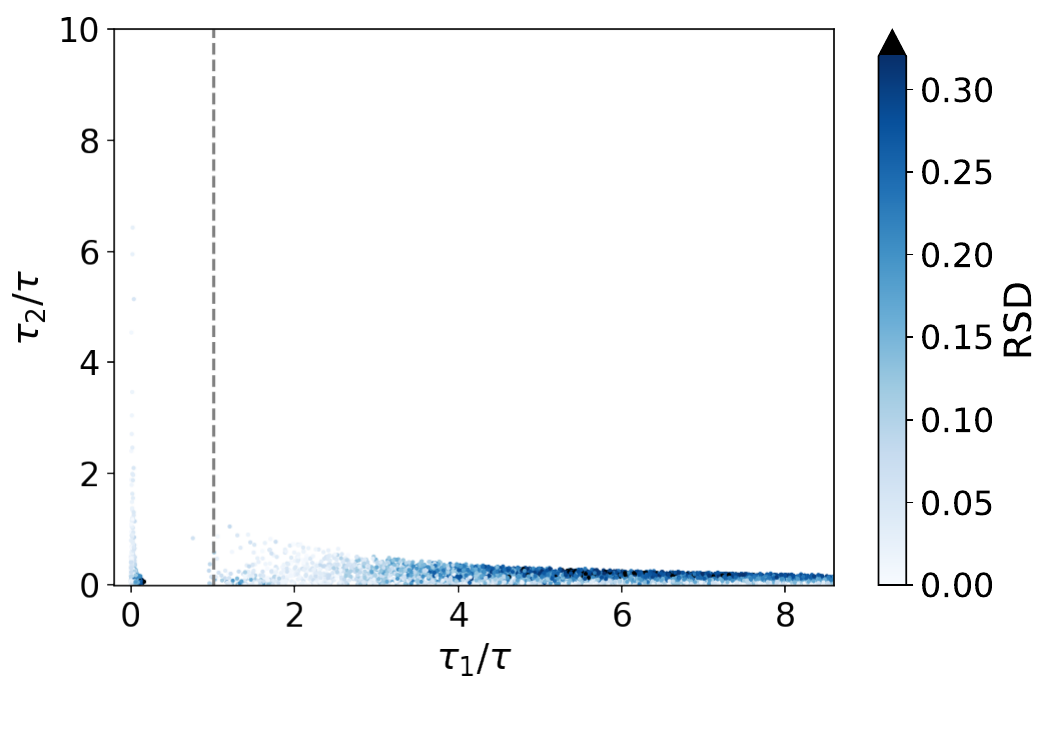}
    \caption{The full size representation of Fig.~\ref{fig_2_random} (c). }
    \label{Fig_2_fulllll}
\end{figure}

\begin{table}[h]
    \centering
\begin{tabular}{c c c c c c}
     \hline \hline \noalign{\vskip 0.3ex}
           &$\alpha_s$ [$\frac{\textrm{fs}}{\textrm{mm}}$] &~$\alpha_i$ [$\frac{\textrm{fs}}{\textrm{mm}}]$&~$\beta_p$ [$\frac{\textrm{fs}\textsuperscript{2}}{\textrm{mm}}]$&~$\beta_s$ [$\frac{\textrm{fs}\textsuperscript{2}}{\textrm{mm}}]$&~$\beta_i$ [$\frac{\textrm{fs}\textsuperscript{2}}{\textrm{mm}}]$\\ [0.4ex]
           \hline 
     Ex1 &   29.15&  -33.79 &  -338 &   131 &-265\\
     Ex2 &   -5.49&  -5.89 &  -11 &   -150 &320\\
     \hline \hline
    \end{tabular}
    \caption{Dispersion parameters for two examples in Fig.~\ref{fig_2_random}.}
    \label{tab:placeholder}
\end{table}

\subsection{Example without a shift }\label{without_shift}

\begin{figure}[h]
    \includegraphics[width=0.5\linewidth]{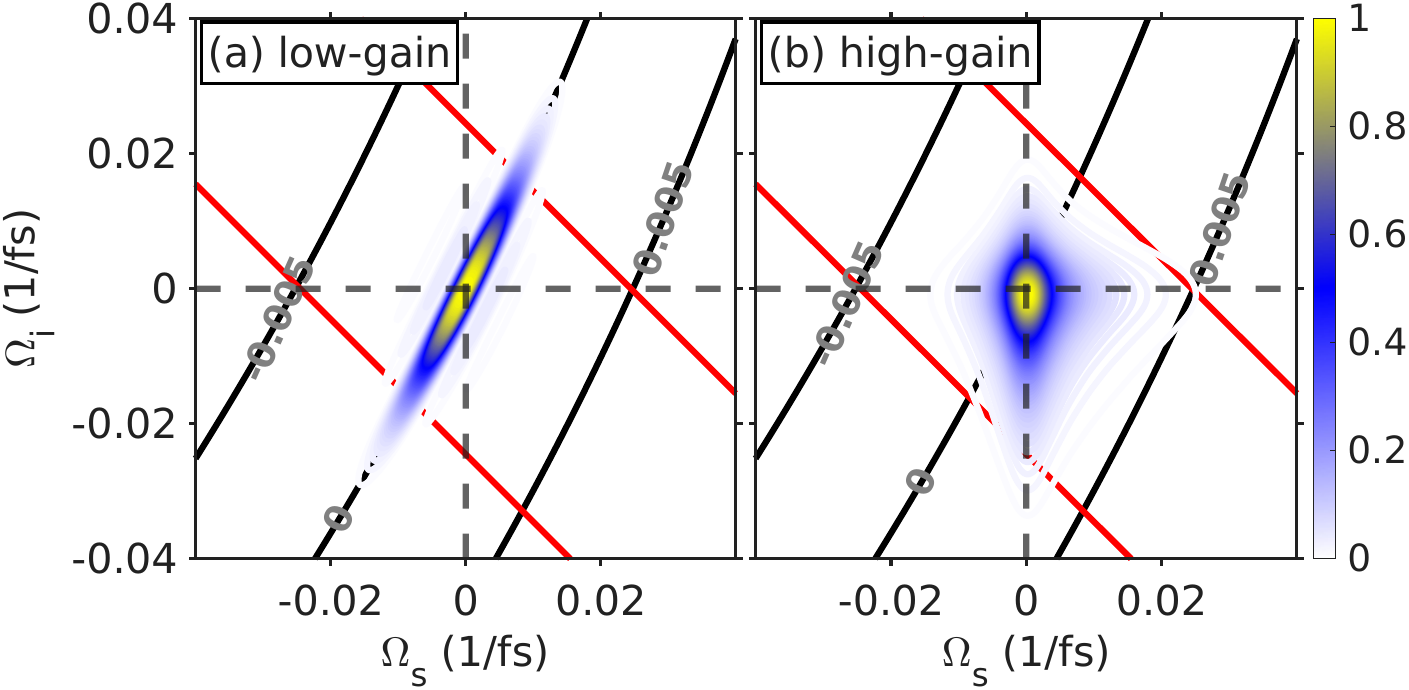}
    \caption{Normalized JSI for (a) low-gain ($N \approx 10^{-3}$) and (b) high-gain ($N\approx10^{5}$) regimes for a waveguide with dispersion parameters presented in the Table~\ref{tab:without_shift}. The frequency shift of the signal and idler distributions is not observed in the high-gain regime due to the almost perpendicular orientation of the PM0 curve to the pump diagonal.}
    \label{no shift}
\end{figure}

Here, we illustrate that the gain-induced non-degeneracy can be absent if the $\Delta k =0$ isoline (PM0 curve) is oriented almost perpendicular to the pump diagonal. For this example, the waveguide dispersion includes $\Delta k_1$ and $\Delta k_2$, which are given in TABLE~\ref{tab:without_shift}. The JSIs in the low- and high-gain regimes are presented in Fig.~\ref{no shift}. 
\begin{table}[h]
    \centering
\begin{tabular}{cccccc}
           \hline \hline \noalign{\vskip 0.3ex}
           &$\alpha_s$ [$\frac{\textrm{fs}}{\textrm{mm}}$] &~$\alpha_i$ [$\frac{\textrm{fs}}{\textrm{mm}}]$&~$\beta_p$ [$\frac{\textrm{fs}\textsuperscript{2}}{\textrm{mm}}]$&~$\beta_s$ [$\frac{\textrm{fs}\textsuperscript{2}}{\textrm{mm}}]$&~$\beta_i$ [$\frac{\textrm{fs}\textsuperscript{2}}{\textrm{mm}}]$\\ [0.4ex]
           \hline 
     No shift WG &   200&  100 &  300 &   150 &100\\
                \hline \hline
    \end{tabular}
    \caption{Dispersion parameters for the waveguide presented in Fig.~\ref{no shift}.}
    \label{tab:without_shift}
\end{table}

\subsection{Signal/idler spectra evolution}\label{evo_section}
\begin{figure}[h]
    \includegraphics[width=0.7\linewidth]{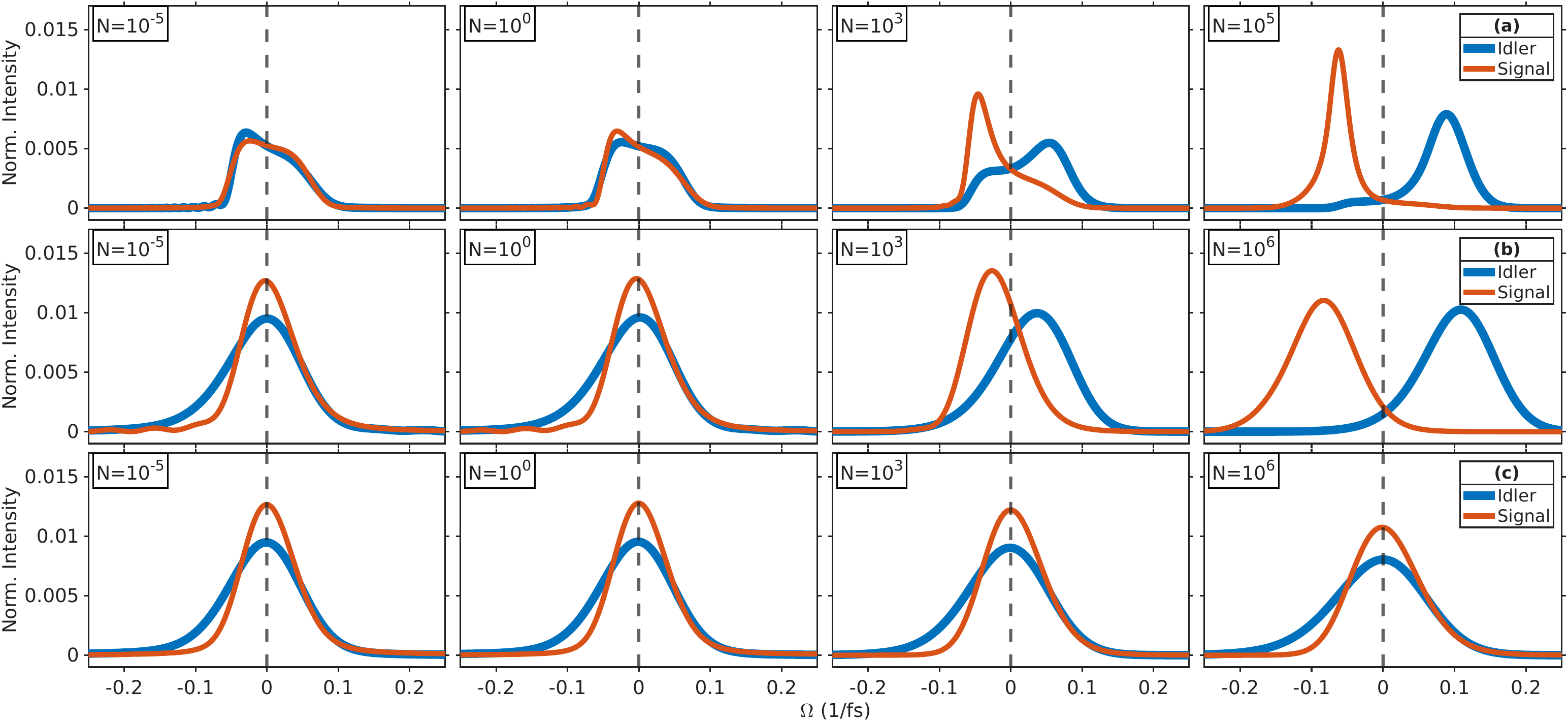}
    \caption{Evolution of normalized spectra of signal (blue) and idler (orange) fields for (a) randomly selected WG (the star in Fig.~\ref{fig_2_random} (c)), (b) WG2, and (c) WG1  with increasing the number of generated photons N.
    }
    \label{evol}
\end{figure}

The gain-evolution of the normalized signal and idler spectra for WG1 and WG2 from the main text, and a random waveguide  (shown as a star in Fig.\ref{fig_2_random} (c)) with the parameters given in the table~\ref{Table:random WG} is presented in Fig.~\ref{evol}. In the low-gain regime, the spectra are highly overlapping and demonstrate the degenerate PDC regime. With increasing the parametric gain (the number of generated photons), the spectra of WG2 and the random waveguide become shifted until their full separation in the high-gain regime. This is not the case for WG1.
One can notice that for a strong second-order dispersion, a complicated modification of the signal and idler spectra can occur with an increase in the gain (see panel (a)). In comparison, for the small second-order dispersion, a well-known spectral broadening takes place (see panel (c)). 
\begin{table}[h]
    \centering
\begin{tabular}{cccccc}
           \hline \hline \noalign{\vskip 0.3ex}
           &$\alpha_s$ [$\frac{\textrm{fs}}{\textrm{mm}}$] &~$\alpha_i$ [$\frac{\textrm{fs}}{\textrm{mm}}]$&~$\beta_p$ [$\frac{\textrm{fs}\textsuperscript{2}}{\textrm{mm}}]$&~$\beta_s$ [$\frac{\textrm{fs}\textsuperscript{2}}{\textrm{mm}}]$&~$\beta_i$ [$\frac{\textrm{fs}\textsuperscript{2}}{\textrm{mm}}]$\\ [0.4ex]
           \hline 
    Random case &   -33.9&  -34.3 &  144 &   -137 &-278\\
               \hline \hline
    \end{tabular}
    \caption{Dispersion parameters for the randomly chosen waveguide shown as a star in Fig.\ref{fig_2_random} (c), the gain-evolution of the spectra for this waveguide is presented in panel (a) of Fig~\ref{evol}.}
    \label{Table:random WG}
\end{table}

\subsection{Spatially-averaged approximate model}\label{AM}

One of the ways to introduce the spatially-averaged model is the replacement of the function $e^{i\Delta k z}$ in Eq.~\eqref{eq_generator_full} with its spatially-averaged form, namely $e^{i\Delta k z} \rightarrow \frac{1}{L}\int_0^L dz e^{i\Delta k z}$. With this replacement, the $z$-dependent coupling matrix $J(\omega,\omega^\prime, z)$ is replaced by its averaged $z$-independent version $\tilde J_L(\omega,\omega^\prime)$, known as the two-photon amplitude (TPA):
\begin{equation}
    \tilde J_L(\omega,\omega^\prime) = S(\omega + \omega^\prime) \mathrm{sinc}\left(\frac{\Delta k(\omega,\omega^\prime) L}{2}\right)e^{\frac{i\Delta k(\omega,\omega^\prime) L}{2}}.   
    \label{eq_TPA}
\end{equation}

This replacement is equivalent to the first-order Magnus approximation~\cite{Horoshko_2019} of the evolution operator $\hat{\mathcal{U}}(z=L) \approx \hat{\mathcal{U}}^{(1)}(L)$, which has a form
\begin{equation}
    \hat{\mathcal{U}}^{(1)}(L) = \exp \left( i \int_0^L dz^\prime \hat{G}(z^\prime)  \right) = \exp \left( i \Gamma L \left[ \iint d\omega d\omega^\prime \tilde J_L(\omega,\omega^\prime) \hat{a}^\dagger(\omega) \hat{b}^\dagger(\omega^\prime)  + h.c.\right] \right).
\end{equation}
Note that this operator does not include the spatial-ordering, meaning that the causality in a corresponding solution is omitted. This approximation results in the spatially-averaged Heisenberg equations: 

\begin{equation}\label{eq_cide_averaged}
\begin{split}
    \frac{d\hat{a}(\omega,z)}{dz}&= i \: \Gamma \int d\omega^\prime \tilde J_L(\omega,\omega^\prime)  b^\dagger(\omega^\prime,z), 
    \\
    \frac{d\hat{b}(\omega,z)}{dz}&= i \: \Gamma \int d\omega^\prime \tilde J_L(\omega,\omega^\prime) a^\dagger(\omega^\prime,z).
    \end{split}
\end{equation}

Note that, although the averaged spatial equations must be solved in the interval from $0$ to $L$, the length $L$ is additionally present as a parameter on the right-hand side of the equations. This means that changing the crystal length in the spatially averaged model requires changing the coupling matrix $\tilde J_L(\omega,\omega^\prime)$.

The solution to the spatially-averaged Heisenberg equations then reads $\hat{a}(\omega,L) = [\hat{\mathcal{U}}^{(1)}(L)]^\dagger \hat{a}(\omega) \ \hat{\mathcal{U}}^{(1)}(L)$, where $\hat{a}(\omega)\equiv\hat{a}(\omega,0)$, and is explicitly given by the following Bogoliubov transformation ~\cite{PhysRevA.97.053827}

\begin{equation}
\begin{split}
\hat{a}^{out}(\omega,L)=\hat{a}^{in}(\omega,0)+\sum_{k}\psi_{k}(\omega)(\cosh(r\sqrt{\lambda_k})-1)A_{k}+\sum_{k}\psi_{k}(\omega)\sinh(r\sqrt{\lambda_k})B_{k}^{\dagger},\\
\hat{b}^{out}(\omega',L)=\hat{b}^{in}(\omega',0)+\sum_{k}\phi_{k}(\omega')(\cosh(r\sqrt{\lambda_k})-1)B_{k}+\sum_{k}\phi_{k}(\omega')\sinh(r\sqrt{\lambda_k})A_{k}^{\dagger},\\
    \end{split}
 \end{equation} 
where $r=\Gamma L$ and the Schmidt decomposition of the TPA is given as $J_L(\omega,\omega^\prime)=\sum_k \sqrt{\lambda_k} \psi_k(\omega) \phi_k(\omega')$. Here $\lambda_k$ is the eigenvalues and $\psi_k(\omega), \phi_k(\omega') $ are the eigenfunctions.    The broadband operators are defined as $A^\dagger(\omega)=\int d\omega \psi(\omega)a^\dagger(\omega)$ and $B^\dagger(\omega')=\int d\omega' \phi(\omega')b^\dagger(\omega')$.

\subsubsection{JSI from average model}
\begin{figure}[h]
    \includegraphics[width=0.7\linewidth]{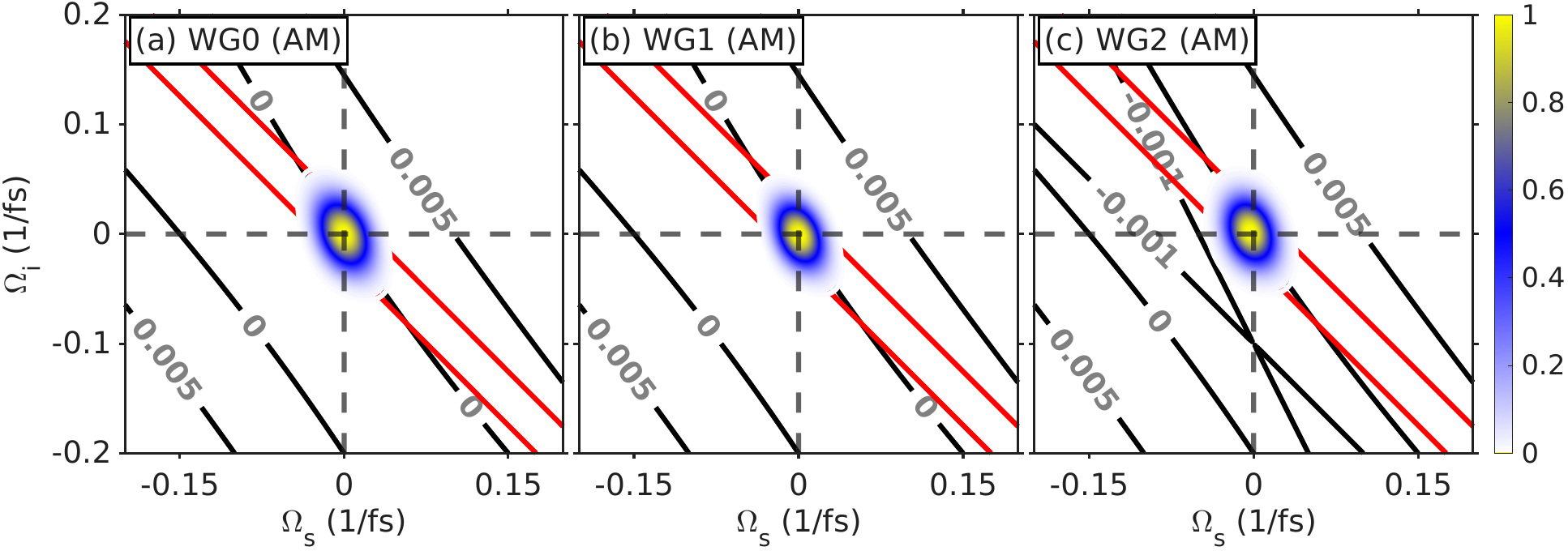}
    \caption{
    Normalized JSIs obtained using the spatially-averaged model in the high-gain regime (N $\approx 10^5$) for (a) WG0, (b) WG1, and (c) WG2.
    }
    \label{fig_JSI_average}
\end{figure}

 Due to neglecting the spatial-ordering effect, the spatially-averaged approximation fails to reproduce the gain-induced frequency shift for all of the waveguides. The normalized JSIs for WG0, WG1, and WG2 obtained using this approximation  are given in
 Fig.~\ref{fig_JSI_average}. Indeed, the averaged model describes correctly the low-gain PDC ($\Gamma L \ll 1$), where the PDC state is approximated by the bi-photon state  $\Psi \propto \ket{0} + \Gamma L \iint d\omega d\omega^\prime \tilde J_L(\omega,\omega^\prime) \hat{a}^\dagger(\omega) \hat{b}^\dagger(\omega^\prime)\ket{0}$. In this case  $\mathrm{JSI}(\omega, \omega^\prime) \propto |J_L(\omega,\omega^\prime)|^2$. For the high-gain PDC, the spatially-averaged model does not provide an accurate description of the generated light, which comes from neglecting the causality (spatial-ordering) in the PDC process. Therefore, the averaged model does not reveal the gain-induced non-degeneracy in WG2 (see Fig.~\ref{fig_JSI_average}) and shows the narrowing of the PDC spectrum incorrectly with increasing gain (see Fig.~\ref{fig_broadening_spectrum}), which additionally emphasizes that the spatial-ordering must be taken into account for describing the high-gain PDC.

\subsection{PPKTP waveguide}
\label{PPKTP}
\begin{table}[h]
    \centering
\begin{tabular}{cccccc}
           \hline \hline \noalign{\vskip 0.3ex}
           &$\alpha_s$ [$\frac{\textrm{fs}}{\textrm{mm}}$] &~$\alpha_i$ [$\frac{\textrm{fs}}{\textrm{mm}}]$&~$\beta_p$ [$\frac{\textrm{fs}\textsuperscript{2}}{\textrm{mm}}]$&~$\beta_s$ [$\frac{\textrm{fs}\textsuperscript{2}}{\textrm{mm}}]$&~$\beta_i$ [$\frac{\textrm{fs}\textsuperscript{2}}{\textrm{mm}}]$\\ [0.4ex]
           \hline 
    PPKTP&  516.6&  221.0 &  292.3 &   30.9 &59.3\\
               \hline \hline
    \end{tabular}
    \caption{Dispersion parameters of the considered PPKTP waveguide.}
    \label{Table:KTP}
\end{table}
\begin{figure}[b]
    \includegraphics[width=0.5\linewidth]{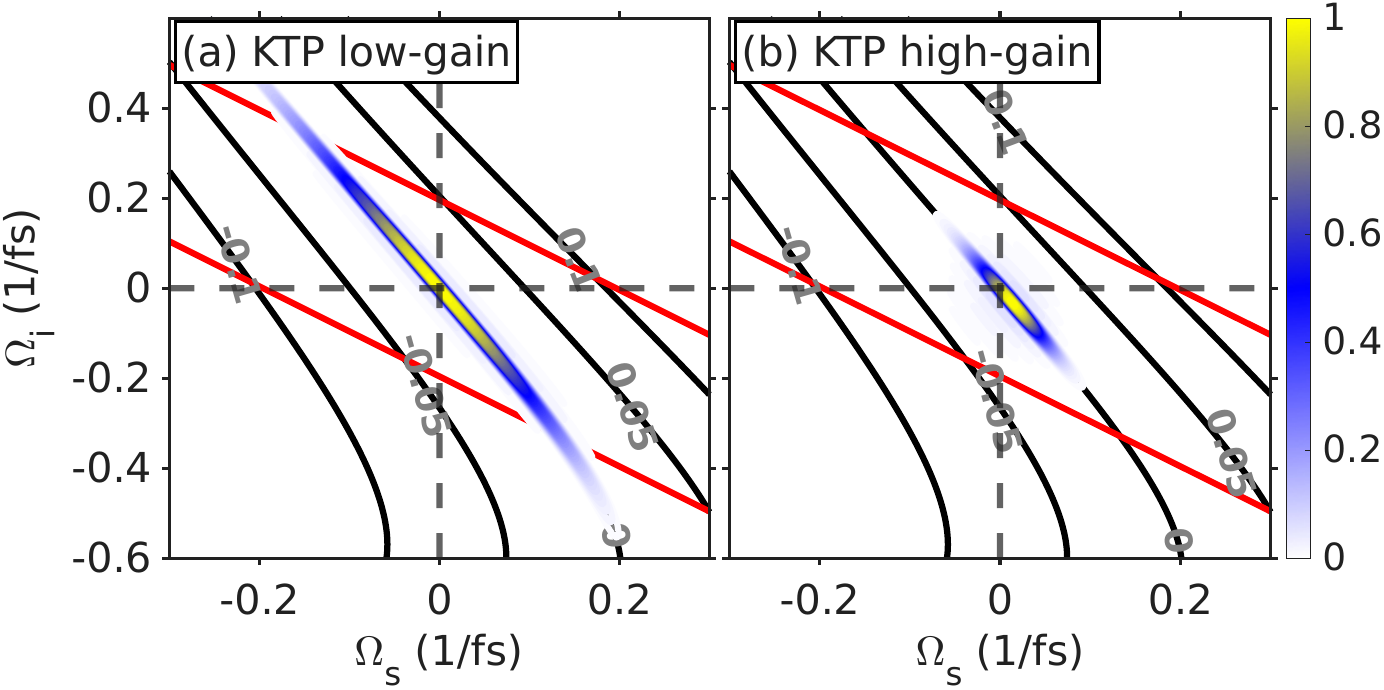}
    \caption{Normalized JSI obtained by using the spatially-averaged model for a 1~mm PPKTP waveguide in a case of (a) low-gain ($N=6 \times10^{-5}$) and (b) high-gain ($N=1.6\times10^{6}$) regimes, the pump pulse duration is 10~fs.}
    \label{KTP_approx}
\end{figure}

To observe the gain-induced non-degeneracy for a real material, we consider the PPKTP waveguide with the poling period of $10.8 \mu m$ to close the $\gamma \rightarrow (\beta, \gamma) $ phase-matching synchronism. The length of the waveguide was fixed as $L=1$ mm, the pump pulse duration was chosen to be $\tau=10$~fs, and the Sellmeier expressions for the refractive indices were chosen to be close to the bulk material \cite{Kato:02}
\begin{equation}
\begin{split}
n^2(\beta) = 3.45018 + \frac{0.04341}{\lambda^2 - 0.04597} + \frac{16.98825}{\lambda^2 - 39.43799},\\
n^2(\gamma) = 4.59423 + \frac{0.06206}{\lambda^2 - 0.04763} + \frac{110.80672}{\lambda^2 - 86.12171}.
\end{split}
\end{equation}

For this case, substituting $\Omega_s = -\Omega_i = 1/\tau$, we can roughly estimate $\Delta k_1^{\mathrm{est}} =(\alpha_s-\alpha_i)/ \tau= 295.6/ \tau \ \mathrm{mm}^{-1}$, $\beta_p/\tau^2=292.3/\tau^2 \mathrm{mm}^{-1}$ 
where $\tau$ is in [fs], see TABLE~\ref{Table:KTP}. In Fig.~\ref{KTP_approx} we present the low- and high-gain JSI for the considered PPKTP waveguide using the spatially-averaged model.

\end{document}